\documentstyle[twocolumn,aps,epsf]{revtex}
\renewcommand{\vec}[1]{\mbox{\boldmath$#1$}}
\newcommand{\tvec}[2]{\left( \begin{array}{c}
   \!\! \mbox{$#1$} \!\! \\ \!\! \mbox{$#2$} \!\! \end{array} \right)}
\begin{document}
\twocolumn[\hsize\textwidth\columnwidth\hsize\csname
@twocolumnfalse\endcsname
%
\title{Andreev Reflection in Narrow Ferromagnet/Superconductor
  Point Contacts} 
\author{K. Kikuchi, H. Imamura$^{*}$, S. Takahashi and S. Maekawa}
  \address{Institute for Materials Research, Tohoku University, Sendai
  980-8577, Japan \newline
  $^{*}$Graduate School of Information Sciences, Tohoku University,
  Sendai 980-8579, Japan}
 \maketitle
%
%
\begin{abstract}
 The Andreev reflection of narrow ferromagnet/superconductor point
 contacts is theoretically studied.  We show that the conductance
 quantization depends on whether the contact region is superconducting or
 ferromagnetic as well as on the strength of the
 exchange field in the ferromagnet.  The Andreev reflection of the
 ferromagnetic contact is more suppressed than that of the
 superconducting contact.
 We also show that the conductance-voltage curve has a bump at zero
 bias-voltage if there is no interfacial-scattering.  
 On the contrary, the conductance-voltage curve shows a
 dip if the system has
 an interfacial-scattering. 
\end{abstract} 
\vskip1pc]%
%
The spin-dependent transport through a magnetic nano-structures is of
current interest both in fundamental physics and application to
spin-electronics\cite{printz1995}.  The magnetic quantum point contact
is one of these magnetic nano-structures, which exhibit rich and
elegant physics\cite{dejong1995,bruno1999,imamura2000} and have
potential applications such as magnetic recording devices
\cite{ono1999,garcia1999,tatara1999}.  Recently much attention has
been focused on magnetic point contacts with superconducting electrode
since it is shown that the spin polarization of the conduction
electrons can be measured by using the
Andreev reflection\cite{soulen1998,soulen1999,upadhyay1998,strijkers2001,ji2001}.

De Jong and Beenakker have shown that the Andreev reflection
in a ferromagnet(F)/superconductor(S) point contact is strongly
suppressed as the Fermi surface polarization is
increased\cite{dejong1995}.  The Fermi surface polarization, $P$, is defined
as $P=(N_{\uparrow}-N_{\downarrow})/(N_{\uparrow}+N_{\downarrow})$,
where $N_{\uparrow(\downarrow)}$is the number of transmitting channels in
the spin-up(spin-down) band.  If $N_{\downarrow} <
N_{\uparrow}$ then only $2N_{\downarrow} =
N_{\downarrow}+N_{\uparrow}(N_{\downarrow}/N_{\uparrow}$)
channels are available for the Andreev
reflection and  the corresponding conductance is $G=4 N_{\downarrow}
e^{2}/h$.  Therefore, the Andreev reflection is
strongly suppressed if $P$ is large.  
The suppression of the Andreev reflection in
F/S point contacts has been observed experimentally by 
several groups\cite{soulen1998,soulen1999,upadhyay1998,strijkers2001,ji2001}.
Their results are
qualitatively well explained by de Jong and Beenakker's theory. 

Now we believe that the narrow F/S point
contact where the conductance is quantized is within the current
nano-technology.  It is then needed to study
the spin-dependent transport through narrow F/S point contacts
taking account of the geometry of contact and the mixing of
channels.  
In this paper, we theoretically study the spin-dependent transport
through narrow F/S point contacts, where the
Andreev reflection plays an important role.
We show that the conductance quantization depends on whether the
contact region
is superconducting or ferromagnetic as well as on
the strength of the exchange field.
The Andreev reflection of the ferromagnetic
contact is more suppressed than that for the superconducting contact.
For the ferromagnetic contact, the width of the contact where the new
transmitting channel opens increases with increasing the exchange field.
We also show that the conductance-voltage curve
shows a bump at zero bias voltage if the 
system has no interfacial-scattering between
the ferromagnet and superconductor.
On the contrary, the conductance-voltage curve shows a dip for 
the contact with an interfacial-scattering.

\begin{figure}
    \epsfxsize=\columnwidth 
    \centerline{\hbox{
        \epsffile{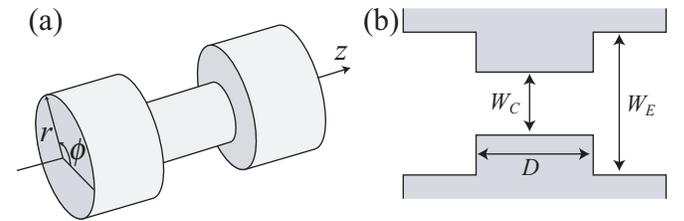}}}
    \caption{(a) The geometry of the point contact.
      The point contact is represented by the
      coaxial cylinders. (b) The cross-section along the $z$-direction.  
    The length and width of the contact region are $D$ and $W_{C}$,
      respectively.  The width of the electrodes is $W_{E}$. }
  \label{fig:model}
\end{figure}

%
We consider the system consisting of three cylinders as shown in
Figs. \ref{fig:model}(a) and  \ref{fig:model}(b).
The left and right electrodes with a diameter $W_{E}$ are connected by
the contact with a diameter $W_{C} (< W_{E})$ and a length $D$.
We employ the simple Stoner model with exchange field $h$ for 
the ferromagnet.

The system we consider is described by the following Bolgoliubov-de
Gennes (BdG) equation:
\begin{equation}
  \left(\!\!\!
    \begin{array}{cc}
      H_{0}(\vec{r}) \!-\! h(z)\sigma       &  \Delta(T)        \\
      \Delta^{\ast}(T)   &  \!-\!H_{0}(\vec{r})\! - \!h(z)\sigma  
    \end{array}
 \! \!\! \right)
  \Psi_{\sigma}(\vec{r})
  \!=\! 
  E
  \Psi_{\sigma}(\vec{r}),
  \label{eq:BdG}
\end{equation}
where $H_{0}(\vec{r})\equiv -(\hbar^{2}/2m)\nabla^{2} + V(\vec{r})
-\mu_{F}$ is the single-particle Hamiltonian with the constriction
potential $V(\vec{r})$, $E$ is the quasiparticle energy measured from
the Fermi energy $\mu_{F}$, $h(z)$ is the exchange field, $\Delta(T)$
is the superconducting energy gap and $\sigma=+(-)$ for the
spin-up(spin-down) band.

We assume that effective mass of an electron $m$ is the same both for
ferromagnet and superconductor.  The exchange field is $h(z)=0$ for
the superconductor and $h(z)=h$ for the ferromagnet.  The constriction
potential is $V(\vec{r})=\infty$ outside the constriction represented
by the shadow in Fig\ref{fig:model}(b).  Inside the constriction,
$V(\vec{r}) = (\hbar^{2}k_{F} Z/m) \delta(z - z_{0})$, where $Z$
represents the interfacial-scattering potential at position $z_{0}$.
We assume that the superconducting energy gap $\Delta(T)$ is constant
and neglect the proximity effects\cite{strijkers2001}.

The stationary scattering state is
obtained by connecting wave functions at the boundary of
contact\cite{furusaki1991,furusaki1992}.
Since the system has cylindrical symmetry, 
wave functions $\Psi_{\sigma}$ can be written as
\begin{equation}
  \Psi_{\sigma}(\vec{r})=  \sum_{n,l} 
  N_{\sigma n}^{l}
  \psi_{\sigma
  l}^{n}(z) J_{n}(\frac{2\gamma_{nl}}{W_{E(C)}}r) e^{i n \phi}
  \label{eq:wf}
\end{equation}
where $\gamma_{nl}$ is the $l$th zero of the Bessel function $J_{n}(r)$,
$N_{\sigma n}^{l}$ is the normalization constant\cite{bogachek1991}.
The set of quantum numbers $(n,l,\sigma)$ defines the channel.
Substituting Eq.(\ref{eq:wf}) into Eq.(\ref{eq:BdG}),
we obtain the following BdG equation for $\psi^{n}_{\sigma l}(z)$:
\begin{equation}
  \left(\!\! \begin{array}{cc}
      H_{0}(z)\! - \!h(z)\sigma       &  \Delta(T)        \\
      \Delta^{\ast}(T)   &  \!-H_{0}(z)\! -\! h(z)\sigma  
    \end{array}
   \!\! \right )
  \psi^{n}_{\sigma l}(z)
   \!= \! E
  \psi^{n}_{\sigma l}(z),
  \label{eq:BdGz}
\end{equation}
where $H_{0}=({\hbar^{2}}/{2m}) \{ {d^{2}}/{dz^{2}} -
({2\gamma_{nl}}/{W(z)})^{2} \}+ (\hbar^{2}k_{F}
Z/m)\delta(z - z_{0})$.

Let us consider the F/S/S system
consisting of ferromagnetic and superconducting electrodes connected
by the superconducting contact.
Assuming that the electron in channel $(n,l,\sigma)$ is incident from the
left electrode, 
the wavefunction $\Psi_{\sigma l}^{n}(z)$ is written as
\begin{eqnarray}
  && \Psi_{\sigma l}^{n}(z)= \nonumber\\
  &&  \left\{
    \begin{array}{l}
      e^{iq_{\sigma l}^{n+} z} 
      \!\tvec{1}{0} 
      J_{n}(\frac{2 \gamma_{n l}}{W_{E}} r) e^{i n \phi}
      + \displaystyle  \sum_{s=1}^{M_{E}}
      \left\{
          a_{\sigma ls}^{n}e^{iq_{\bar{\sigma} s}^{n-} z}
          \!\!\tvec{0}{1}\right. \\
        \left. \mbox{\hspace{1em}} 
          +  b_{\sigma ls}^{n}e^{-iq_{\sigma s}^{n+} z}
          \!\!\tvec{1}{0}
        \right\} \times
        J_{n}(\frac{2 \gamma_{n s}}{W_{E}} r) e^{i n \phi}\\
        \mbox{\hspace{14em}}
        ({\rm for\ \ }z < -\frac{D}{2}) \\ \\
        \displaystyle \sum_{s=1}^{M_{C}}
        \left[
        \left\{
        \alpha_{\sigma ls}^{n}e^{ip_{s}^{n+} z}
        +\beta_{\sigma ls}^{n}e^{-ip_{s}^{n+} z}
        \right\} 
        \!\!\tvec{u_{0}}{v_{0}}
        \right.
        \\
        \left.
        \mbox{\hspace{1em}}  +\{\xi_{\sigma ls}^{n}e^{ip_{s}^{n-} z}
          +\eta_{\sigma ls}^{n}e^{-ip_{s}^{n-} z}
          \} 
          \!\!\tvec{v_{0}}{u_{0}}
          \right]\\
        \mbox{\hspace{1em}}\times
        J_{n}(\frac{2 \gamma_{n s}}{W_{C}} r) e^{i n \phi}\\
        \mbox{\hspace{12em}} ({\rm for \ \ }-\frac{D}{2} < z <
      \frac{D}{2}) \\ \\
      \displaystyle \sum_{s=1}^{M_{E}}
      \left[c_{\sigma ls}^{n}e^{ik_{s}^{n+} z}
        \!\!\tvec{u_{0}}{v_{0}}
        +d_{\sigma ls}^{n}e^{-ik_{s}^{n-} z}
        \!\!\tvec{v_{0}}{u_{0}}
        \right]\\
        \mbox{\hspace{1em}}\times
        J_{n}(\frac{2 \gamma_{n s}}{W_{E}} r) e^{i n \phi}\\
        \mbox{\hspace{15em}}({\rm for \ \ }\frac{D}{2} < z),
      \end{array} \right. 
  \label{eq:pc_wavefunction}
\end{eqnarray}
where $\bar{\sigma}=-\sigma$, $u_{0}^{2} = 1-v_{0}^{2} =
 ( E+\sqrt{E^{2}-\Delta(T)^{2}})/2$ and
 $a_{\sigma lm}^{n}$, $b_{\sigma lm}^{n}$,
 $c_{\sigma lm}^{n}$, $d_{\sigma lm}^{n}$,
 $\alpha_{\sigma lm}^{n}$, $\beta_{\sigma lm}^{n}$,
 $\xi_{\sigma lm}^{n}$, $\eta_{\sigma lm}^{n}$
 are coefficients to be solved.
Here the wave vectors below the superconducting energy gap $(E \le
 \Delta(T))$ are given by 
\begin{eqnarray}
  q_{\sigma l}^{n +}&=&\sqrt{\frac{2m}{\hbar^2} \left( 
                       \mu_{F} + h\sigma + E \right)
                        -(\frac{2\gamma_{nl}}{W_E})^2} \\
  q_{\sigma l}^{n -}&=&\left(
                       \sqrt{\frac{2m}{\hbar^2} \left( 
                       \mu_{F}- h\sigma- E \right)
                        -(\frac{2\gamma_{nl}}{W_E})^2}
                      \right)^{*}\\
  p_{l}^{n\pm} &=& \sqrt{ \frac{2m}{\hbar^2}
    \left(
      \mu_{F}\pm\sqrt{E^2-\Delta(T)^2} 
    \right)
    -(
    \frac{2\gamma_{nl}}{W_C})^2}\\
  k_{l}^{n\pm}&=&\sqrt{ \frac{2m}{\hbar^2} \left(
                \mu_{F}\pm\sqrt{E^2-\Delta(T)^2} \right)
                -(\frac{2\gamma_{nl}}{W_E})^2}.
              \label{eq:wvec}
\end{eqnarray}
The wave vectors above the superconducting energy gap $(E >
\Delta(T))$ are the same as those given by Eq.(\ref{eq:wvec})
except for the wave vector $k_{l}^{n-}$ in the right electrode.
In order to deal with the evanessent waves, we write the wave
vector $k_{l}^{n-}$ as
\begin{eqnarray}
  k_{l}^{n-}&=&\left( \sqrt{ \frac{2m}{\hbar^2} \left(
                       \mu_{F}-\sqrt{E^2-\Delta(T)^2} \right)
                       -(\frac{2\gamma_{nl}}{W_E})^2} \right)^{\ast}.
\end{eqnarray}
The number of channels in the electrode and contact is truncated by
the cutoff constant $M_E$ and $M_C$, respectively.  The cutoff
constants are taken to be large enough to express the stational
scattering state.

For the F/F/S system, where the contact is ferromagnetic, the
wavefunction inside the contact in Eq.(\ref{eq:pc_wavefunction}) is
rewritten as
\begin{eqnarray}
  &&\displaystyle \sum_{s=1}^{M_{C}}
  \left[
    \left\{
      \alpha_{\sigma ls}^{n}e^{ip_{\sigma s}^{n+} z}
      +\beta_{\sigma ls}^{n}e^{-ip_{\sigma s}^{n+} z}
    \right\} 
    \!\!\tvec{1}{0}
  \right.\nonumber \\
  && 
  \left.
    +\{\xi_{\sigma ls}^{n}e^{ip_{\bar{\sigma}s}^{n-} z}
    +\eta_{\sigma ls}^{n}e^{-ip_{\bar{\sigma}s}^{n-} z}
    \} 
    \!\!\tvec{0}{1}
  \right]
  \times
  J_{n}(\frac{2 \gamma_{n s}}{W_{C}} r) e^{i n \phi}
\end{eqnarray}
where the wave vectors $p_{\sigma l}^{n\pm}$ are given by
\begin{equation}
  p_{\sigma l}^{n\pm} = \sqrt{ \frac{2m}{\hbar^2}
    \left(
      \mu_{F}\pm h \sigma \pm E
    \right)
    -(
    \frac{2\gamma_{nl}}{W_C})^2}.
\end{equation}
    
Unknown coefficients
 $a_{\sigma lm}^{n}$, $b_{\sigma lm}^{n}$,
 $c_{\sigma lm}^{n}$, $d_{\sigma lm}^{n}$,
 $\alpha_{\sigma lm}^{n}$, $\beta_{\sigma lm}^{n}$,
 $\xi_{\sigma lm}^{n}$ and  $\eta_{\sigma lm}^{n}$
are obtained by matching the slope and value of wave functions
at the boundary of contact $z=\pm \frac{D}{2}$\cite{furusaki1991,furusaki1992}.
Following Blonder, Tinkham and Klapwijk\cite{blonder1982}, 
probabilities for the Andreev reflection, $A_{ls}^{n}$, and
ordinary reflection, $B_{ls}^{n}$, are given by 
\begin{eqnarray}
  A_{\sigma ls}^{n}(E) &=& \left| \frac{q_{\bar{\sigma} s}^{n-}J_{n-1}^{2}(\gamma_{ns})}
    {q_{\sigma l}^{n+}J_{n-1}^{2}(\gamma_{nl})} \right|
  a_{\sigma ls}^{n\ast} a_{\sigma ls}^{n} 
  \\
  B_{\sigma ls}^{n}(E) &=&\left| \frac{q_{\sigma s}^{n+}J_{n-1}^{2}(\gamma_{ns})}
    {q_{\sigma l}^{n+}J_{n-1}^{2}(\gamma_{nl})} \right|
  b_{\sigma ls}^{n\ast} b_{\sigma ls}^{n}.
\end{eqnarray}

The differential
conductance is expressed by using the 
probabilities $A_{ls}^{n}$ and $B_{ls}^{n}$ as
\begin{eqnarray}
  G &=& \frac{e}{h} \sum_{\sigma=\uparrow,\downarrow}
  \sum_{n,l}\int_{-\infty}^{\infty}
  \frac{d}{dV} f(E - eV,T)\nonumber\\
  && \times\left\{1+\sum_{s}A_{\sigma
      ls}^{n}(E)-\sum_{s}B_{\sigma ls}^{n}(E) \right\} dE,
\end{eqnarray}
where $f(E,T)$ is the Fermi distribution function.

In our numerical calculation, the cutoff constants $M_C$ and $M_E$ are
taken to be three times as many as the number of open channels.
The number of open channels are determined by the condition that
${\rm Re}[(2m/\hbar^2)(\mu_{F}-\sqrt{E^2-\Delta(T)^2})
-(2\gamma_{nl}/W_C(E))^2 ] > 0$.
The diameter of electrodes and the length of contact are taken to
be $W_{E}=60.8/k_{F}$ and $D=5.0/k_{F}$, respectively.  
The superconducting energy gap is assumed to be $\Delta(0)/\mu_{F}=1.5\times
10^{-5}$, which is of the same order of that for Al\cite{ashcroft}.
The position of the interfacial-scattering is located at $z_{0} = -D/2
(+D/2)$ for the F/S/S (F/F/S) system.

\begin{figure}
    \epsfxsize=\columnwidth 
    \centerline{\hbox{
        \epsffile{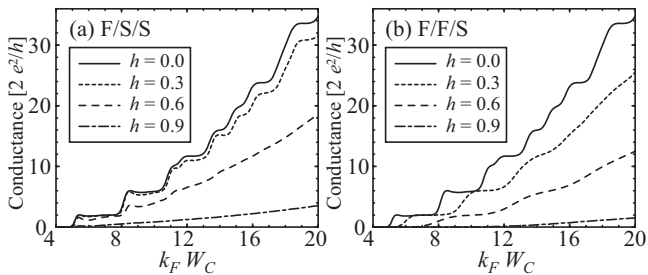}}}
    \caption{(a) The zero-bias conductance for the F/S/S point
      contact.  The temperature $T$ is assumed to be zero and no
      interfacial-scattering is considered.  The horizontal axis represents the
      width of the contact $W_{C}$ multiplied by the Fermi-wave
      number $k_{F}$. 
      The solid, dotted, dashed and dot-dashed
      lines represent the conductance curves for $h_{0}=0, 0. 3, 0.6$
      and $0.9$, respectively.
      (b) The same plot for the F/F/S point contact.} 
  \label{fig:quantization}
\end{figure}

In Fig. \ref{fig:quantization} (a) the zero-bias conductance of the F/S/S system
is plotted against the width of contact $W_{C}$.  In the adiabatic
picture, the number of 
transmitting channels are determined by the condition that ${\rm
  Re}[(2m/\hbar^2)(\mu_{F}-\sqrt{E^2-\Delta(T)^2})
-(2\gamma_{nl}/W_E)^2 ] > 0$ and does not depend on the
strength of the exchange field in the ferromagnet electrode.  
However, as the exchange field increases, the conductance is
suppressed due to the mismatch of the Fermi wavelength
as shown in Fig. \ref{fig:quantization}(a).

The conductance of the F/F/S system also decreases as the exchange
field increases as shown in Fig. \ref{fig:quantization}(b). 
Note that the width $W_{C}$ at which the new
transmitting channel opens increases with increasing the exchange
field, $h$.
The shift of the conductance steps can be explained as follows.
In the ferromagnetic contact, electrons in the spin-up and spin-down
bands feel the different exchange field $-h$ and $h$,
respectively.  Therefore, the number of transmitting spin-down
channels $N_{\downarrow}$ is smaller than that of transmitting spin-down
channels $N_{\uparrow}$.   As pointed out by de Jong and Beenakker.
The number of channels contributing to the Andreev reflection is
restricted by $N_{\downarrow}$.
In the adiabatic picture\cite{yacoby1990}, 
$N_{\downarrow}$ is determined by the condition
that $(2m/\hbar^2)( \mu_{F} - E - h) - (2\gamma_{nl}/W_C)^2  > 0$.  
Therefore, the conductance steps shift leftward with increasing
the exchange field.

In the narrow F/F/S system, the suppression of the Andreev reflection
discussed by de Jong and Beenakker appears as the shift of 
the width of the contact $W_{C}$ at which the new transmitting channel
opens as shown in Fig. \ref{fig:quantization}(b).
Even when the contact is superconducting(F/S/S), the conductance is suppressed
due to the mismatch of the Fermi wavelength as shown in
Fig. \ref{fig:quantization}(a).  However, the magnitude of suppression
is smaller than that for the F/F/S system.

\begin{figure}
    \epsfxsize=\columnwidth 
    \centerline{\hbox{
        \epsffile{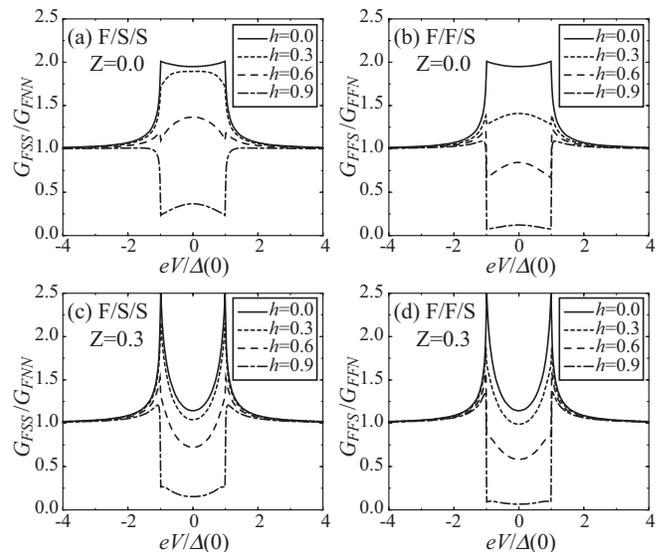}}}
    \caption{The conductance for the F/S/S and F/F/S point
      contacts is plotted against the bias voltage at $T=0$.
      The width of the contact is taken to be $k_{F}W_{C}=18.5$.
      The strength of the interfacial-scattering $Z$ is assumed to be
      zero for panels (a) and (b), and $Z=0.3$ for panels (c) and (d).
      The horizontal axis represents the applied bias voltage
      normalized by the superconducting gap $\Delta(0)$ at $T=0$.
      The solid, dotted, dashed and dot-dashed
      lines represent the conductance curves for the $h_{0}=0, 0. 3, 0.6$
      and $0.9$, respectively.} 
  \label{fig:zero-t}
\end{figure}

Figures \ref{fig:zero-t}(a) and \ref{fig:zero-t}(c) show the
conductance(G)-voltage(V) curves
for the F/S/S system and Figs.\ref{fig:zero-t}(b) and
\ref{fig:zero-t}(d) show those for the F/F/S system.
The conductance is normalized by that for the F/normal-metal(N)
contact. One can see that the conductance for the F/F/S system is smaller
than that for F/S/S system since the number of transmitting channels
are limited by the Andreev reflection.

In order to investigate the effect of the interfacial-scattering, we plot
the G-V curve for the interfacial-scattering of $Z=0.3$ in
Figs.\ref{fig:zero-t}(c) and \ref{fig:zero-t}(d).  Without the
interfacial scattering, $Z=0$, the G-V curve has a
bump at zero bias voltage as shown in Figs.\ref{fig:zero-t}(a)
and \ref{fig:zero-t}(b).  However, the G-V curve for  $Z=0.3$
shows dip as shown in Figs.\ref{fig:zero-t}(c)
and \ref{fig:zero-t}(d).  These dip and bump in the G-V
curve are similar to the energy dependence of conductance for
F/S tunnel junctions discussed by \v{Z}uti\'{c} and Valls\cite{Zutic2000}.

Comparing our theory with the experimental results of Soulen 
{\it et  al.}\cite{soulen1998,soulen1999}, we conclude that in their
experiments there is no interfacial-scattering  in the contacts with
Ni, NiFe and Co film, whereas the interfacial-scattering exists in
the contacts with NiMnSb, La$_{0.7}$Sr$_{0.3}$Mn$_{3}$ and CrO$_{2}$
film.

Soulen {\it et al.}\cite{soulen1998} have also studied the conductance
of two different systems with the same material: a
sharpened Ta point placed in contact with a single crystal Fe thin
film, and a sharpened Fe point placed in contact with a
polycrystalline Ta foil.  They found the difference in
the G-V curve near zero bias voltage and concluded that this difference
is due to varying amounts of interfacial-scattering, $Z$.
However, we propose that such difference in the G-V curve can occur if
the material of the contact is different: one system has the
superconducting contact and the other system has the ferromagnetic contact.
We also insist that the spin-polarization obtained by analyzing the
experimental date\cite{upadhyay1998,strijkers2001,ji2001} depends
strongly on whether the contact region is ferromagnetic or
superconducting.

\begin{figure}
    \epsfxsize=\columnwidth 
    \centerline{\hbox{
        \epsffile{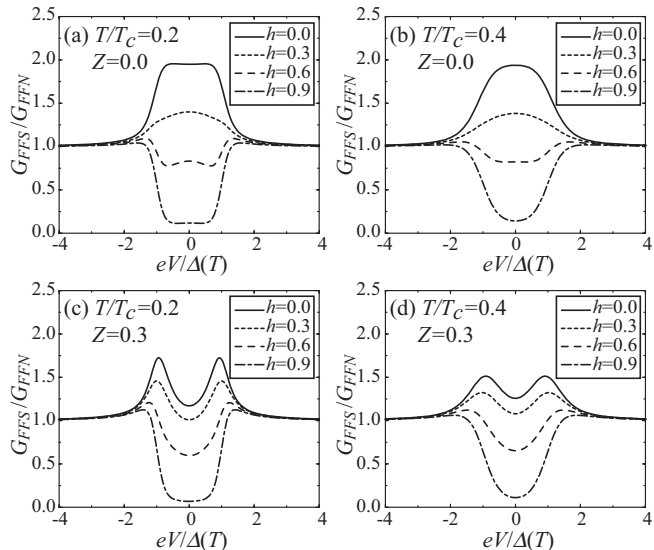}}}
    \caption{The conductance for the F/F/S point
      contacts is plotted against the bias voltage.
      The width of the contact is taken to be $k_{F}W_{C}=18.5$.
      The strength of the interfacial-scattering $Z$ is assumed to be
      zero for panels (a) and (b), and $Z=0.3$ for panels (c) and (d).
      The temperature is assumed to be $T/T_{c}=0.2$ for panels (a) and (c),
      and $T/T_{c}=0.4$ for panels (b) and (d).  The horizontal axis
      represents the applied bias voltage
      normalized by the superconducting gap $\Delta(T)$.
      The solid, dotted, dashed and dot-dashed
      lines represent the conductance curves for the $h_{0}=0, 0.3, 0.6$
      and $0.9$, respectively.} 
  \label{fig:finite-t}
\end{figure}

In Figs.\ref{fig:finite-t}(a)-\ref{fig:finite-t}(d), we show the
G-V curve of the F/F/S system at $T/T_{c}=0.2$ and 0.4.
The horizontal axis is normalized by the superconducting energy gap
$\Delta(T)$.  The temperature dependencies of the G-V curves for the
F/S/S system are not shown since they are similar to those for the
F/F/S system shown in
Figs.\ref{fig:finite-t}(a)-\ref{fig:finite-t}(d).
One can clearly see that the difference of the curvature of the G-V
curve due to the interfacial-scattering survives even at $T/T_{c}=0.2$.
As the the temperature is increased, the difference in the curvature
of the G-V curve for large exchange filed $h=0.6$ and 0.9 smears out.
For small exchange filed $h=0.0$ and 0.3, however, the curvature of
the G-V curve due to the interfacial-scattering survives even at
$T/T_{c}=0.4$.

In conclusion, we theoretically study the Andreev reflection of
narrow F/S point contacts.
We show that the conductance depends on whether the
contact is superconducting or ferromagnetic as well as on
the strength of the exchange field in the ferromagnet.
The conductance of the ferromagnetic contact is more
suppressed than that of the superconducting contact.
For the ferromagnetic contact, the width of the contact where the new
transmitting channel opens increases with increasing the exchange
field.  We also show that the conductance-voltage curve has a bump
at zero bias-voltage if there is no interfacial-scattering between
the ferromagnet and superconductor.  On the contrary, if the contact
has an interfacial-scattering, the conductance-voltage curve has a
dip at the zero bias-voltage.

This work is supported by a Grand-in-Aid for Scientific Research from MEXT.
H.I. is supported by MEXT, Grant-in-Aid for
Encouragement of Young Scientists,13740197, 2001.
S.M. acknowledges support of the Humboldt Foundation.

\end{document}